# Sub-exponential Upper Bound for #XSAT of some CNF Classes

Bernd R. Schuh

Dr. Bernd Schuh, D-50968 Köln, Germany; bernd.schuh@netcologne.de

keywords: complexity, satisfiability, SAT, XSAT, #XSAT, propositional logic

**Abstract.**

We derive a worst upper bound on the number of models for exact satisfiability (XSAT) of arbitrary CNF formulas *F*. The bound can be calculated solely from the distribution of positive and negated literals in the formula. For certain subsets of CNF instances the new bound can be computed in sub-exponential time, namely in at most $O(n^{\sqrt{n}})$, where n is the number of variables of *F*. A wider class of SAT problems beyond XSAT is defined to which the method can be extended.

**Introduction.**

Hard instances of SAT and SAT variants are characterized by exponential decision times. Therefore methods to determine whether a specific instance can be expected to be hard or not are of interest. We will introduce such a method in the following and apply it to some SAT variants. A frequently investigated SAT variant is XSAT, short for exact satisfiability. XSAT is the problem of deciding whether assignments exist which satisfy exactly one literal in each clause. The study of this class of problems is interesting, because it is known to be NP-complete on conjunctive normal form formulas (CNF). The natural quantity to test the conditions imposed by XSAT is the number of true (i.e. satisfied) literals $\sigma_F$, because in XSAT the number of true literals necessarily equals the number of clauses, $\sigma_F = m$. Not all assignments fulfilling this condition need to be XSAT models, i.e. XSAT satisfying assignments. But their number can serve as an upper bound for the effort to solve the problem.

A simplified version of the method and a generalized XSAT problem has been used in a former paper to identify SAT problems which can be decided in polynomial time [1]. Also the class of monotone *l*-regular exact linear CNF formulas, shorthand $XLCNF_+^l$, was investigated with the help of this approach [2,3], supplementing the extensive studies by Porschen et.al. on linear and exact linear CNFs [4,5]. As a result, any XSAT instance from this class was shown to be either XSAT-unsatisfiable or its satisfiability could be decided in sub-exponential time of order $O(n^{\sqrt{n}})$.

In this paper conditions on general CNF formulas are formulated which lead to sub-exponential decidability with respect to XSAT and related SAT variants.

We use notations and definitions as introduced in [2]. Additionally, with $l_{js}$ we denote the literal of variety *s* in clause *j*, i.e. $l_{js}$ is either $a_s$ or the negated $\overline{a_s}$ occurring in clause *j*. Furthermore we introduce the three quantities $p_s, p_{s+}, p_{s-}$ for the total number of literals of atom $a_s$, the number of positive and negated literals of variety *s*, respectively. *N*=total number of literals in *F*, suppressing a subscript *F* if the context is clear.

The new feature here is to split the formula *F* into blocks of variables with the same absolute difference in the occurrence of negated and positive literals. As a result the probability distribution of $\sigma_F$ with respect to assignment space and thus the number of assignments which lead to a given value of $\sigma_F$ can be calculated from the basic parameters $p_{s+}, p_{s-}$.

Results are presented as a couple of lemmata and theorems in the next section. Their significance is discussed in a concluding section.

**Results.**

A central expression for the number of true literals is derived in the first

<u>Lemma 1</u>: Let *y* denote an arbitrary assignment $y : V(F) \to \{1,0\}$ and $y_s := y(a_s)$. Let $\sigma_F(y) = \sum_{j,s} y(l_{js})$ denote the number of true literals under this assignment. Then

(1) $\quad \sigma_F(y) = \sum_{s=1}^{n} \left( p_{s+} y_s + p_{s-}(1 - y_s) \right).$

Proof: If $y_s = 1$ all positive literals are true and contribute a 1 to the sum each. For $y_s = 0$ the other literals do. □

Now we split the variables into non-intersecting blocks defined by equal difference of occurrence of positive and negated variables. Then $\sigma_F$ can be split into a sum over blocks of variables, as shown in

Lemma 2:

Define a variable block $V_b(F) := \{a_s \in V(F) : |p_{s+} - p_{s-}| = b\}$ for each $b \in \mathbb{N}_{|F|} \cup \{0\}$. Denote by $B_F$ the set of integers $b$ for which $V_b(F) \neq \emptyset$, and by $I_b := \{s \in \{1,...,n\} : |p_{s+} - p_{s-}| = b\}$ the index set corresponding to $V_b(F)$. We set $n_b := |I_b|$ for short, and $N_b := \sum_{s \in I_b} p_s$ for the number of literals of variable block $b$. Then:

(2) $\quad \sigma_F(y) = \sum_{b \in B_F} \sigma_F^{(b)}(y) \text{ with } \sigma_F^{(b)} = \frac{1}{2} N_b + \frac{1}{2} b \sum_{s \in I_b} sign(p_{s+} - p_{s-})(2y_s - 1)$

Proof: Straightforward calculation, starting with equation (1). □

Lemma 3: $\sigma_{F\min} := \min_y \sigma_F(y) = \frac{1}{2}(N - \sum_{b \in B_F} bn_b)$ and $\sigma_{F\max} := \max_y \sigma_F(y) = \frac{1}{2}(N + \sum_{b \in B_F} bn_b)$ are the minimum and maximum values of $\sigma_F$ with respect to all possible assignments. They are attained with assignments $2y_s^{\min/\max} - 1 = \mp sgn(p_{s+} - p_{s-})$, respectively.

Proof: Follows immediately from equation (2). □

Lemma 4: If $x_s := 2y_s - 1$ changes sign (i.e. the variable $a_s$ is flipped), $\sigma_F^{(b)}$ changes by an amount $b$.

Proof: Obvious from equation (2). The statement holds for $b = 0$, too. □

**Theorem** 1: Define a quantity $\Delta\sigma^{(b)}$ for $b \neq 0$ by $\Delta\sigma^{(b)}(y) := \frac{\sigma_F^{(b)}(y)}{b} - \frac{1}{2}\left(\frac{N_b}{b} - n_b\right)$. Then $\Delta\sigma^{(b)}$ can take values from $\{0, 1, ..., n_b\}$ with frequency $\binom{n_b}{\Delta\sigma^{(b)}}$.

Proof: From lemma 4 and equation (2) it is clear that $\sigma_F^{(b)}/b$ changes by $\pm 1$ if any $y_s$ ($s \in I_b$) changes from 1 to 0 or vice versa. From lemma 3 one concludes that the minimum possible value of $\Delta\sigma^{(b)}(y)$ is 0, and its maximum value $n_b$. Now for the frequency. To change the value of $\sigma_F$ from its minimum value to $\sigma_{F\min} + \Delta\sigma^{(b)}$ one must flip $\Delta\sigma^{(b)}$ many of the $n_b$ variables of $V_b(F)$. Since there are $\binom{n_b}{\Delta\sigma^{(b)}}$ possible ways to do that, the theorem is proved.

For a more formal proof one calculates the characteristic function of each $\Delta\sigma^{(b)}(y)$

$$\chi_{\Delta\sigma^{(b)}}(\alpha) := E[\exp(\alpha\Delta\sigma^{(b)})] = 2^{-n_b} \sum_{\{y_s \in \{1,0\}; s \in I_b\}} \exp(\alpha(\Delta\sigma^{(b)}(y)))$$

and gets $\chi_{\Delta\sigma^{(b)}}(\alpha) = 2^{-n_b}(1+\exp(\alpha))^{n_b}$. Thus the $\Delta\sigma^{(b)}$ are independent random variables with binomial distribution. □

Now we are in a position to formulate the central theorem.

**Theorem** 2: Let *F* be a CNF formula, *y* an assignment on the variables of *F*, and $\sigma_F(y)$ the corresponding number of true literals in *F* when the assignment is applied. Let $\sigma_0$ be some integer between the minimum and maximum values of $\sigma_F$. Write $A_F(\sigma_0)$ for the set of assignments y which satisfy $\sigma_F(y) = \sigma_0$, i.e. $A_F(\sigma_0) := \{y : \sigma_F(y) = \sigma_0\}$. Then the number of elements of $A_F(\sigma_0)$ is given by

(3) $\qquad |A_F(\sigma_0)| = 2^{n_0} \sum_{(w_b)} \left( \prod_{b \in B_F} \binom{n_b}{w_b} \right)$,

where the sum runs over all integer tuples $(w_b) \in \underset{b \in B_F}{\times} (\mathbb{N}_{n_b} \cup \{0\})$ which are solutions of

(4) $\qquad \sigma_0 - \sigma_{F\min} = \sum_{b \in B_F} bw_b$.

$B_F$ and $n_b$ are defined in lemma 2.

Proof: From the definition of $\Delta\sigma^{(b)}$ in theorem 1 one can express $\sigma_F$ as

$$\sigma_F = \sum_{b \in B_F} \sigma_F^{(b)}(y) = \sum_{b \in B_F} b\Delta\sigma^{(b)} + \frac{1}{2}(N - \sum_{b \in B_F} bn_b) \text{ or, according to lemma 3: } \sigma_F = \sum_{b \in B_F} b\Delta\sigma^{(b)} + \sigma_{F\min}.$$

Thus the condition $\sigma_F(y) = \sigma_0$ is equivalent to $\sigma_0 - \sigma_{F\min} = \sum_{b \in B_F} b\Delta\sigma^{(b)}(y)$. Since for each $b \in B_F$ there are $\binom{n_b}{\Delta\sigma^{(b)}}$ assignments leading to he same $\Delta\sigma^{(b)}$ one gets the product over $b \in B_F$ for each $|B_F|$-tuple of integers $\Delta\sigma^{(b)}$ fulfilling this equation. The factor $2^{n_0}$ arises because variables with $p_{s+} = p_{s-}$ contribute the same number of true literals for $y_s = 1$ as well as $y_s = 0$. □

From these statements one can immediately derive an algorithm to decide certain SAT variants. To this end we use a SAT variant termed PART-SAT, which was introduced in [3].

Definition: For any CNF formula F with m clauses let $m = \mu_0 + \mu_1 + ... + \mu_k$ be a partition of m with $k \leq \max_{C \in F} |C|$. Define "PART-SAT" as the following decision problem: does a truth assignment exist such that $\mu_\alpha$ many clauses contain exactly $\alpha$ true literals each.

As an example set $\mu_\alpha = 0$ for all $\alpha$ except $\alpha = 1$, i.e. $\mu_1 = m$. Then PART-SAT coincides with XSAT. To specify the particular PART-SAT problem we write $\{\mu_0, \mu_1, ..., \mu_k\}$ - SAT. E.g. XSAT for 3-CNF, usually termed "1-in-3SAT" in this notation reads $\{0, m, 0, 0\}$ - SAT.

Lemma 5: For a $\{\mu_0, \mu_1, ..., \mu_k\}$ - SAT instance F every satisfying assignment y fulfills $\sigma_F(y) = \sum_{\alpha=1}^{k} \alpha \mu_\alpha$.

The statement follows immediately from the definition of PART-SAT.
Assignments which fulfill the equation in lemma 5 will be called *pseudomodels* (with respect to PART-SAT). Those pseudomodels which are fulfilling assignments of the PART-SAT problem will be called models (with respect to PART-SAT), as usual.

**Theorem 3**: Given a PART-SAT problem $\{\mu_0, \mu_1, ..., \mu_k\}$ and an instance F to be tested. Let $A_F(\sum_{\alpha=1}^{k} \alpha \mu_\alpha)$ be the set of pseudomodels. Then F can be decided in time less than $O\left(\left|A_F(\sum_{\alpha=1}^{k} \alpha \mu_\alpha)\right|\right)$ up to polynomial time factors. The bound is given by $\left|A_F(\sum_{\alpha=1}^{k} \alpha \mu_\alpha)\right| = 2^{n_0} \sum_{(w_b)} \left(\prod_{b \in B_F} \binom{n_b}{w_b}\right)$ where now the integers $w_b$ have to be determined from $\sum_{b \in B_F} b w_b = \sum_\alpha \alpha \mu_\alpha - \sigma_{F \min}$ with boundary conditions $0 \leq w_b \leq n_b$.

Proof: The second part of the theorem is an immediate consequence of the definition of pseudomodels, lemma 5 and theorem 2. For the first part one has to name a decision algorithm which has a computation time of at most of order $\left|A_F(\sum_{\alpha=1}^{k} \alpha \mu_\alpha)\right|$ up to polynomial time corrections. An appropiate algorithm works as follows: $\sigma_{F \min}$ can be determined by a simple

counting process: either by the formula given in lemma 3 or $\sigma_{F\min} = \sum_{s=1}^{n} \min(p_{s+}, p_{s-})$. The corresponding assignment (or assignments, if $n_0 \neq 0$) has been given in lemma 3 already. Starting from this or these minimal assignments one can build up for each tuple of solutions $(w_b)$ the corresponding pseudomodels by flipping as many variables as indicated by the tuples. This process is $O(\prod_{b \in B_F} \binom{n_b}{w_b})$ by construction. The elements of $A_F(\sum_{\alpha=1}^{k} \alpha \mu_\alpha)$ given, one can now test each of them for the PART-SAT condition, which is possible in p.t.. Also the solution of the equation $\sum_{b \in B_F} b w_b = \sum_\alpha \alpha \mu_\alpha - \sigma_{F\min}$ cannot take more steps than is inherent in the binomial factors already, because a brute force method would be to probe all possibilities allowed by the boundary conditions, i.e. $\prod_{b \in B_F} n_b$ many. □

We give a simple example to illustrate the use of the theorem. Let $F = \{(a_1, a_2, a_3), (a_1, a_2, \overline{a_5}), (\overline{a_1}, a_2, \overline{a_3}, a_4), (\overline{a_3}, \overline{a_4}, \overline{a_5})\}$ in an obvious notation. We ask for exact satisfiability, i.e. an assignment which evaluates exactly one literal to true in each clause. First determine $B_F$ and $\sigma_{F\min}$: $B_F = \{0, 1, 2, 3\}$ with multiplicities $n_0 = n_2 = n_3 = 1$ and $n_1 = 2$. $\sigma_{F\min} = 3$. According to theorem 3 we first need to solve $\sum_{b \in B_F} b w_b = w_1 + 2w_2 + 3w_3 = m - \sigma_{F\min} = 1$. The only solution is $w_2 = w_3 = 0$ and $w_1 = 1$. Thus the number of pseudomodels is $2 \cdot 2 \cdot 1 \cdot 1 = 4$. To find the corresponding assignments one starts from the "groundstate" corresponding to $\sigma_{F\min}$. It reads $(y_1, y_2, y_3, y_4, y_5) = (0, 0, 1, *, 1)$, i.e. there two states (*) corresponding to the two possible values for $y_4$. The solution $w_2 = w_3 = 0$, $w_1 = 1$ means we must not flip the variables with $b=2$ and $b=3$, but switch one of the variables with $b=1$, i.e. either $a_1$ or $a_3$. So the 4 pseudomodels are $(1, 0, 1, *, 1)$ and $(0, 0, 0, *, 1)$ with $* \in \{0, 1\}$. Unfortunately, none of these is an XSAT solution. The algorithm returns the answer "x-UNSAT". Had we asked however for the PART-SAT problem $\{1, 3, 0, 0\} - SAT$, i. e. a solution which leaves exactly one clause unsatisfied and assigns one true literal to each of the remaining clauses, we would have gotten a positive answer.

We close this section by noting that also usual SAT can be formulated as a PART-SAT problem and the algorithm of theorem 3 is applicable. We state the fact as a

Theorem 4: Let $F$ be a CNF SAT instance. Then the number of steps to decide whether $F \in SAT$ is bounded from above by $2^{n_0} \sum_{l=m}^{\sigma_{\max}} \sum_{(w_b(l))} \left( \prod_{b \in B_F} \binom{n_b}{w_b(l)} \right)$ where the $(w_b(l))_{b \in B_F}$ have to be calculated from

$$\sum_{b \in B_F} bw_b(l) = l - \sigma_{F\min}$$ for all integer $l$ with $m \leq l \leq \sigma_{F\max}$.

Proof: For an assignment $y$ which evaluates all clauses of $F$ to true, each clause must contain at least one true literal. Therefore $\sigma_F(y) \geq m$ is a necessary condition. Furthermore, since $\sigma_F(y)$ is bounded from above by $\sigma_{F\max}$ all possible pseudomodels for SAT are among the solutions of one of the equations $\sigma_F(y) = l$ with $m \leq l \leq \sigma_{F\max}$. There can be no double counting because an assignment cannot evaluate the same formula to a different number of true literals. Thus theorem 3 can be applied for each $l$ separately and the results can be added up to the expression in the theorem. □

**Conclusion.**

We have derived bounds on the number of satisfying assignments for SAT (in theorem 4) and SAT variants like PART-SAT (theorem 3), including XSAT and others, and we now discuss whether these bounds can yield useful results. It is not difficult to construct problems, for which the derived bounds definitely grow exponentially with problem size. There are classes of CNF formulas, however, for which the bound does not explode exponentially with problem size, i.e. $\sim \alpha^n$ with some $1 < \alpha \leq 2$, but only underline{exponentially in the square root} of the problem size, i.e. $\sim \beta^{\sqrt{n}}$ with some $1 < \beta$ (which is at most a polynomial in $n$). To illustrate this sub-exponential decidability we consider the bound for #XSAT, i.e. the expression for the number of pseudomodels of an XSAT instance $F$ with $m$ clauses and $n$ variables. One has to solve equations (3) and (4) of theorem 2, setting $\sigma_0 = m$. Some steps in the calculation can be performed in linear or polynomial time, like the determination of $B_F$ or $\sigma_{F\min}$. They just involve basic parameters like $p_{s+}, p_{s-}$ or $\sigma_{F\min} = \sum_s \min(p_{s+}, p_{s-})$. Crucial points are the determination of the $w_b$ and the binomials. This is where – in the limit $n \to \infty$ – exponential behavior may come into play.

As a first example consider the extreme case where there is only one (non-zero) $b$-value: $B_F = \{l\}$, and consequently $n_b = n$. Monotone $l$-regular CNF formulas are examples of this class. In this case equ. (3) reduces to $|A_F(m)| = \binom{n}{w}$, and (4) to $m - \sigma_{F\min} = lw$. As an interesting consequence we see that $F$ is x-unsat if $m - \sigma_{F\min}$ is not a multiple of $l$, because in this case the second equation does not have a solution. The same conclusion holds, of course, whenever $m < \sigma_{F\min}$. If this is not the case and also $m - \sigma_{F\min} = 0 \bmod(l)$ holds then the bound can be calculated from $|A_F(m)| = \binom{n}{(m-\sigma_{F\min})/l}$. This formula yields an interesting result for the subclass of monotone exact linear *l*-regular CNF formulas, as shown in a preceding paper, [2]. In this case one can prove that $(m-\sigma_{F\min})/l = m/l = O(\sqrt{n})$ for large *n*. This leads to a sub-exponential behavior of the binomial of order $O(n^{\sqrt{n}})$. Thus the number of XSAT models of this CNF class is sub-exponentially bounded.

A second example is the other extreme case: formulas where all *b*-values occur just once. Consequently $n_b = 1$ and $w_b \in \{0,1\}$ for each $b \in B_F$. Then equ. (4) amounts to an integer partition of $\rho := m - \sigma_{F\min}$. Since each element of $B_F$ appears just once the number of pseudomodels determined by equ. (3) is smaller than the total number of partitions of the integer $\rho$, $p(\rho)$. Since the partition function $p$ varies as $p(\rho) \sim \frac{1}{4\sqrt{3}\rho}\exp(\pi\sqrt{2\rho/3})$ asymptotically for large $\rho$, it is shown that XSAT problems on this class of formulas can be solved in sub-exponential time, <u>provided</u> that an additional condition $\rho \sim O(n)$ holds. To illustrate that such formulas indeed exist, consider the (artificial) class of instances defined by $F \in CNF$ with $\min\{p_{s+}, p_{s-}\} = p'$; $\max\{p_{s+}, p_{s-}\} = s + p'$; $m = \lambda n$ for all $s \in \{1,2,...,n\}$ and fixed integers $\lambda > p'$. These instances fulfill $B_F = \mathbb{N}_n$, each $b \in B_F$ occurs exactly once, and $m - \sigma_{\min} = \lambda n - np' = (\lambda - p')n = O(n)$. Thus for the XSAT problem "find an assignment with one true literal per clause" the method yields a set of pseudomodels whose cardinality is limited by the partition function $p(m-\sigma_{\min}) \sim O(\frac{1}{\sqrt{n}}\exp\sqrt{n})$.

Irrespective of the relevance of the CNF classes discussed above one can always benefit from the pseudomodel bounds in investigating specific instances. Whether a problem may be solved efficiently or not may vary from instance to instance. So the bounds can be helpful in estimating whether a specific instance is to be considered "hard" or "easy" in terms of exponential decision times. In view of theorem 4 this statement even holds for SAT problems.